\documentclass[aps,onecolumn,superscriptaddress]{revtex4-2}
\usepackage{graphicx}
\usepackage{amsmath,amssymb}
\usepackage{xcolor}
\usepackage{listings}
\usepackage{hyperref}



\hypersetup{
    colorlinks=true,        
    linkcolor=blue,         
    citecolor=blue,         
    filecolor=blue,         
    urlcolor=blue           
}

\begin{document}
\title{Chirally-sensitive optical rectification by isotropic chiral media}

\author{Raju Adhikary}
\affiliation{Department of Physical and Chemical Sciences, University of L'Aquila, Via Vetoio, 67100 L'Aquila, Italy}

\author{Luca Assogna}
\affiliation{Department of Physical and Chemical Sciences, University of L'Aquila, Via Vetoio, 67100 L'Aquila, Italy}

\author{Ambaresh Sahoo}
\affiliation{Department of Physical and Chemical Sciences, University of L'Aquila, Via Vetoio, 67100 L'Aquila, Italy}

\author{Matteo Venturi}
\affiliation{Department of Physical and Chemical Sciences, University of L'Aquila, Via Vetoio, 67100 L'Aquila, Italy}

\author{Andrea De Marcellis}
\affiliation{Department of Physical and Chemical Sciences, University of L'Aquila, Via Vetoio, 67100 L'Aquila, Italy}

\author{Massimiliano Aschi}
\affiliation{Department of Physical and Chemical Sciences, University of L'Aquila, Via Vetoio, 67100 L'Aquila, Italy}

\author{Antonio Mecozzi}
\affiliation{Department of Physical and Chemical Sciences, University of L'Aquila, Via Vetoio, 67100 L'Aquila, Italy}

\author{Jens Biegert}
\affiliation{ICFO - Institut de Ciencies Fotoniques, The Barcelona Institute of Science and Technology, Castelldefels 08860, Barcelona, Spain}
\affiliation{ICREA, Pg. Llu\'is Companys 23, 08010 Barcelona, Spain}

\author{Davide Tedeschi}
\affiliation{Department of Physical and Chemical Sciences, University of L'Aquila, Via Vetoio, 67100 L'Aquila, Italy}

\author{Carino Ferrante}
\affiliation{CNR-SPIN, c/o Dip.to di Scienze Fisiche e Chimiche, Via Vetoio, Coppito (L'Aquila) 67100, Italy}

\author{Andrea Marini}
\affiliation{Department of Physical and Chemical Sciences, University of L'Aquila, Via Vetoio, 67100 L'Aquila, Italy}
\affiliation{CNR-SPIN, c/o Dip.to di Scienze Fisiche e Chimiche, Via Vetoio, Coppito (L'Aquila) 67100, Italy}

\email{andrea.marini@univaq.it}

\begin{abstract}
Chiroptical sensing is central to gain fundamental insight into electronic, vibrational and rotational degrees of freedom of chiral molecules, and is a cornerstone for nanomedicine and drug discovery platforms. Current chiral sensing technologies to assess the enantiomeric imbalance of chiral pharmaceutical compounds are  sensitive to ml volumes but are time-consuming and cannot be integrated on a chip, thus creating a major bottleneck for drug discovery and nanomedicine. Here, we propose a novel chiroptical sensing approach based on optical rectification in a photonic micro-cavity filled by a drug solution with nl volume. We theoretically demonstrate that, upon optical excitation by intense pulsed laser light, such a nonlinear effect produces a chirally-sensitive nV voltage burst at the electrically-gated micro-cavity boundaries, with sign depending solely on the drug enantiomeric imbalance. Our results shed light on the potential of optical rectification as a robust platform for innovative lab-on-a-chip devices enabling chiral sensing with nl sensitivity.
\end{abstract}

\maketitle
\section{Introduction} 

Chiral molecules exist in either of their non-superimposable forms—so-called enantiomers—each exhibiting distinct functional characteristics from their mirror counterparts, which play a crucial role particularly in pharmacodynamics, pharmacokinetics and toxicity \cite{McConathy2003,ChiralDrugs2006,Smith2009}. In turn, efficient chiral sensing technologies are in high demand for precise enantiomeric discrimination in pharmaceutical, drug discovery and drug designing industries. Typically, enantiomeric imbalance in chiral molecular mixtures is assessed by nuclear magnetic resonance \cite{Shundo2009}, X-ray crystallography \cite{Deschamps2010}, gas chromatography \cite{Manoli2013}, capillary electrophoresis \cite{Yu2019}, and high performance liquid chromatography \cite{Okamoto2008}. However, such techniques are not suitable for real-time analysis, lab-on-a-chip
integration, and wider nanomedicine applications. Photonics, e.g., polarimetry \cite{Schreier1995} and electronic/vibrational circular dichroism (CD) \cite{Muller2000,Wesolowski2013}, constitutes a promising platform to overcome such limitations. However, because such techniques exploit magnetic dipole transitions, the associated chiroptical signals are weak and their detection requires ml operation volumes. Enhanced chiroptical sensitivity can be attained locally by the exploitation of superchiral fields \cite{Tang2010, Tang2011}, pushing forward enantiomeric detection limits to the nanoscale thanks to advanced photonic platforms, e.g.,  metasurfaces and nanophotonic structures \cite{Mohammadi2018,Pellegrini2018,Gilroy2019,Im2024}. In particular, plasmon-enhanced CD by metal-based nanostructures provides advanced sensitivity thanks to surface plasmon waves enabling the amplification of light-matter interaction   \cite{Govorov2011,GovorovNanoshell,Nesterov2016,Venturi2023,Raju2025}.   Furthermore, ultrafast and nonlinear (NL) chiroptical spectroscopies \cite{Ayuso2022} enable advanced understanding of chirality at the molecular level, e.g., time-resolved dynamical chiral processes \cite{Beaulieu2018,Neufeld2019}, ultrafast imaging of chiral dynamics \cite{Ayuso2019}, and electronic ring currents \cite{Neufeld2019_2}. However, in spite of such scientific advancements, the development of practical photonic devices with chiroptical sensitivity to nl volumes remains a formidable challenge, hindering disruptive applications in nanomedicine and drug discovery.

\begin{figure*}[t!]
\includegraphics[width=0.99\textwidth]{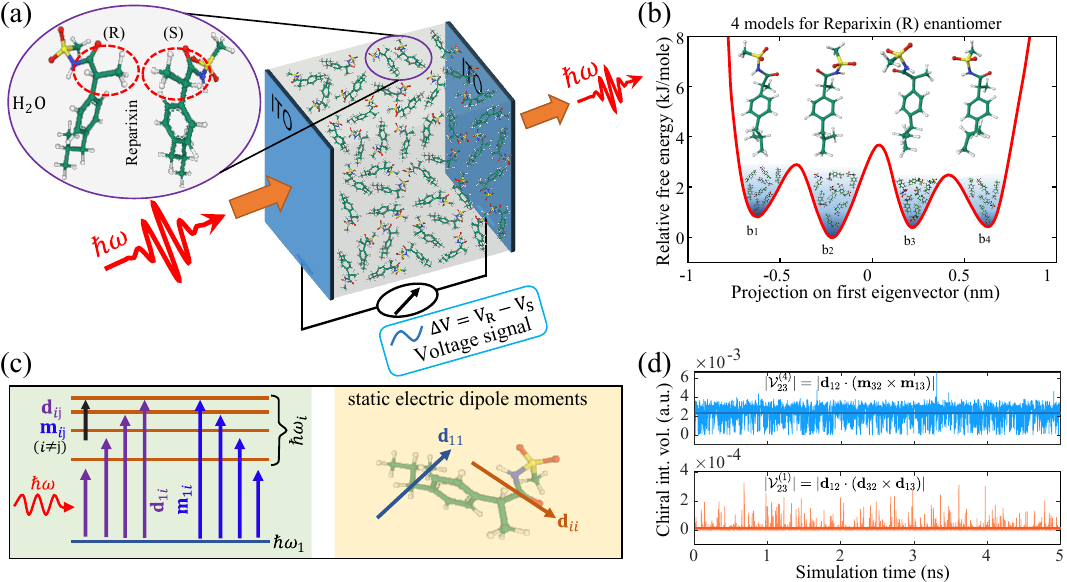}
\caption{(\textbf{a}) Schematic illustration of the considered PMC composed of two ITO electrodes separated by a distance $d\simeq 100 \ \mu{\rm m}$ embedding reparixin dissolved in water. (\textbf{b}) One-dimensional cross-section of the conformational basins of pure S-reparixin enantiomer
dissolved in water calculated by projecting the MD-generated cartesian coordinates of aqueous reparixin onto the first eigenvector of the all-atom covariance matrix, evaluated at temperature ${\rm T} = 298\ {\rm K}$. Note the co-existence of four distinct, statistically relevant thermodynamic conformational basins $b_{\rm k}$, where ${\rm k}=1-4$, in equilibrium. (\textbf{c}) Schematic representation of the electronic structure of reparixin and the considered electronic transitions upon excitation by laser field in the electric/magnetic dipole approximation. (\textbf{d}) Temporal evolution of chiral interaction volumes moduli $|\mathcal{V}_{23}^{(1)}|$ and $|\mathcal{V}_{23}^{(4)}|$ of a single reparixin molecule in the $b_1$ conformation
state calculated at each MD simulation frame (expressed in nanoseconds) employing a cubic simulation box with volume $27\ {\rm nm}^3$,  complemented with PMM calculations \protect \cite{Venturi2023}.}
\label{Fig1}
\end{figure*}

Here we develop an innovative platform to attain enhanced chiroptical sensitivity to nl drug volumes by optical rectification (OR) in photonic microcavities (PMCs) embedding the chiral sample, see Fig.~\ref{Fig1}(a). Such a second-order NL effect produces a quasi-static polarization (QSP) through intense optical excitation of the considered system. While OR typically occurs in non-centrosymmetric media \cite{BoydBook}, we report, for the first time to our knowledge, that isotropic assemblies of randomly oriented chiral molecules produce a chirally-sensitive QSP superimposed to a background QSP, both parallel to the Poynting vector and spin density associated to the pump optical field. For this reason, the electrical detection of the chirally-sensitive signal requires two transparent conducting electrodes perpendicular to the optical pump propagation direction, see Fig.~\ref{Fig1}(a). In our calculations, we consider Indium Tin Oxide (ITO) \cite{Landoni2014} as transparent conductor thanks to its widespread technological exploitation and its biocompatible functionality \cite{Selvakumaran2000}. In order to assess the device functionality on realistic chiral drugs, we consider reparixin (an inhibitor of the CXCR2 function attenuating inflammatory responses \cite{Goriod2007} that has been adopted in clinical trials for the treatment of hospitalized patients with COVID-19 pneumonia \cite{Landoni2022}) dissolved in water with dilute number
molecular density $n_{\rm mol} = n_{\rm mol}^{\rm (R)} + n_{\rm mol}^{\rm (S)} \simeq 10^{-2} \ {\rm nm}^{-3}$ (corresponding to a concentration of 5 ${\rm mg/ml}$), where $n^{\rm (R,S)}_{\rm mol}$ indicate the molecular number densities of R and S enantiomers, respectively, see Fig.~\ref{Fig1}(a). 

We observe that, when pumped by intense laser light pulses with ns time duration and carrier vacuum wavelength $\lambda$, such a chiral mixture produces a GHz voltage signal with nV peak at the ITO electrodes. Remarkably, we observe that the peak voltage is modulated by the enantiomer imbalance density $\Delta n_{\rm mol} = n_{\rm mol}^{(\rm S)} - n_{\rm mol}^{(\rm R)}$, and in turn by measuring such voltage, it is possible to retrieve the enantiomeric excess. Our
results are based on molecular dynamics (MD), time-dependent Density Functional Theory (TD-DFT), and Perturbed Matrix Method (PMM) simulations (accounting for the solvent perturbation) for the calculation of the reparixin electronic structure and of its static and transition electric/magnetic dipole moments \cite{Venturi2023}. All macroscopic optical parameters including second-order NL susceptibilities of the isotropic chiral mixture, are calculated from first principles by perturbatively solving the density matrix equations of electrons in reparixin. This is accomplished by considering the leading electronic transitions producing ultraviolet (UV) absorption peaks and by averaging the obtained hyper-polarizability tensors over the random molecular orientation through the Euler rotation matrix approach. The generated OR voltage signal is calculated through the rigorous solution of vectorial Maxwell's equations in the considered PMC. Our predictions of efficient OR in isotropic chiral drug solutions with sub-nl volume offer novel avenues for integrated chiroptical sensing in lab-on-a-chip devices, which hold great promise for future applications in drug discovery and nanomedicine.

\section {Results}

\subsection{Microscopic Electron Dynamics}

\begin{figure*}[t!]
\includegraphics[width=0.99\textwidth]{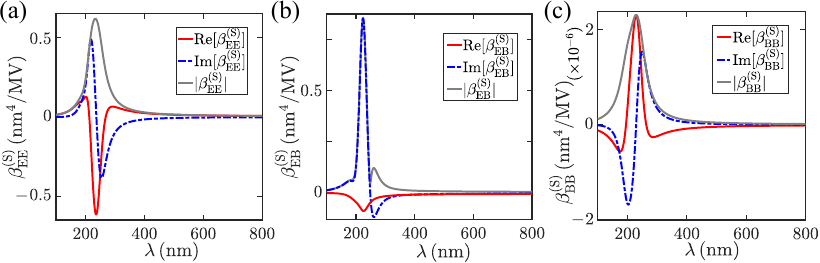}
\caption{Dependence of isotropic hyper-polarizabilities (\textbf{a}) $\beta_{\rm EE}^{\rm (S)}(\lambda)$, (\textbf{b}) $\beta_{\rm EB}^{\rm (S)}(\lambda)$, and (\textbf{c}) $\beta_{\rm BB}^{\rm (S)}(\lambda)$ over the vacuum wavelength $\lambda$ for the S-reparixin enantiomer dissolved in water.}
\label{Fig2}
\end{figure*}

We model the radiation-induced microscopic response of reparixin within the density matrix framework, incorporating quantum molecular observables (QMOs) of the drug solution. In principle, for a flexible solvated molecule like aqueous reparixin, the QMOs need to be treated as true quantum expectation values and determined as averaged statistical-mechanical quantities over a large ensemble of reparixin-water configurations. We adopt previously reported results for the solvated reparixin QMOs \cite{Venturi2023} (i.e., electric and magnetic dipole moments, electronic excitation energies), obtained through a combination of semi-classical MD simulations (for sampling reparixin-water conformations), TD-DFT (to obtain solvent-unperturbed QMOs), and PMM (to account for solvent effects) \cite{Aschi2001, Amadei2009, Carrillo2017}. The MD-trajectory analysis through the Essential Dynamics reveals that reparixin, when dissolved in water, exhibits four statistically relevant conformations corresponding to four equilibrium free energy basins \cite{Venturi2023}. Fig.~\ref{Fig1}(b) represents the one-dimensional cross-section of the relative free-energy basin labeled as $b_{\rm k}$ with $k = 1,2,3,4$, calculated for pure S-reparixin enantiomer dissolved in water. We emphasize that we only consider the electronic states $i= 1 - 5$ that correspond to resonant electronic absorption peaks in the UV. Within the electric/magnetic dipole approximation, the interaction of each reparixin enantiomer specified by the index ${\rm q =  R, S}$, with external radiation is governed by the total Hamiltonian 
\begin{equation} \label{eq1}
    \hat{\mathcal{H}}_{\rm tot}^{(k,{\rm q})}({\bf r},t) = \sum_{i=1}^5 \hbar \omega_i(k) |\phi_i(k,{\rm q}) \rangle \langle \phi_i(k,{\rm q})| - {\bf E} ({\bf r},t) \cdot \hat{\bf d}(k,{\rm q}) - {\bf B} ({\bf r},t) \cdot \hat{\bf m}(k,{\rm q}),
\end{equation}
where $\hbar \omega_i (k)$ is the radiation-unperturbed electronic energy eigenvalues with corresponding eigenstates $|\phi_i(k,{\rm q}) \rangle$, ${\bf E} ({\bf r},t)$ and ${\bf B} ({\bf r},t)$ are the electric and magnetic fields of the external radiation, respectively. In addition, $\hat{\bf d}(k,{\rm q})=\sum_{i,j=1}^5 {\bf d}_{i,j}(k,{\rm q}) |\phi_i(k,{\rm q}) \rangle \langle \phi_j(k,{\rm q})|$ and $\hat{\bf m}(k,{\rm q})=\sum_{i,j=1}^5 {\bf m}_{i,j}(k,{\rm q}) |\phi_i(k,{\rm q}) \rangle \langle \phi_j(k,{\rm q})|$ are the dyadic representations of electric and magnetic electronic dipole operators in the set of unperturbed electronic energy states, respectively. Upon excitation by a driving laser field, reparixin exhibits multilevel electronic transitions from both the ground state $\hbar \omega_1 \rightarrow \hbar \omega_i$ (from the electronic ground state $i=1$ to higher energy states $i =2,3,4,5$) and excited states $\hbar \omega_i \rightarrow \hbar \omega_j$ (between higher energy states $i,j \geq 2$ with $i \neq j$). We schematically illustrate such electron dynamics in Fig.~\ref{Fig1}(c), indicating (left panel) electric ${\bf d}_{1i}(k,{\rm q}), {\bf d}_{ij}(k,{\rm q})$ and magnetic ${\bf m}_{1i}(k,{\rm q}), {\bf d}_{ij}(k,{\rm q})$ transition dipole moments and (right panel) static electric ${\bf d}_{11}(k,{\rm q}), {\bf d}_{ii}(k,{\rm q})$ dipole moments. Within the Born-Oppenheimer approximation, the coexistence of two electrons with opposite spins in the ground state of reparixin leads to vanishing static magnetic dipole moments for every electron eigenstate ${\bf m}_{ii}=0$. Because the electric dipole moments are (purely real) polar vectors ${\bf d}_{i,j}(k,{\rm q}) \in \mathbb{R}^3$ and magnetic dipole moments are (purely imaginary) axial vectors ${\bf m}_{i,j}(k,{\rm q}) \in \mathbb{I}^3$, for opposite enantiomers they transform as ${\bf d}_{i,j}(k,{\rm R}) = \mathcal{R}_{\hat{\bf n}} {\bf d}_{i,j}(k,{\rm S})$ and ${\bf m}_{i,j} (k,{\rm R}) = -\mathcal{R}_{\hat{\bf n}} {\bf m}_{i,j} (k,{\rm S})$, where $\mathcal{R}_{\hat{\bf n}} = {\rm I} - 2{\hat{\bf n}}{\hat{\bf n}}$ is the reflection operator across the broken-symmetry plane perpendicular to the unit vector $\hat{\bf n}$. Applying such transformations, one can evaluate the chirally-sensitive properties of triple products involving ${\bf d}_{i,j}(k,{\rm q})$ and ${\bf m}_{i,j}(k,{\rm q})$, entering second-order NL interactions, see below. Importantly, we observe that while 
\begin{subequations}
\begin{align}
& {\mathcal V}_{1,i}(k,{\rm q}) = [{\bf d}_{i,i}(k,{\rm q}) - {\bf d}_{1,1}(k,{\rm q})] \cdot [{\bf d}_{1,i}(k,{\rm q}) \times {\bf m}_{1,i}(k,{\rm q})], \\
& {\mathcal V}_{i,j}^{(2)}(k,{\rm q}) = {\bf d}_{1,i}(k,{\rm q}) \cdot [{\bf d}_{1,j}(k,{\rm q}) \times {\bf m}_{j,i}(k,{\rm q})], \\
& {\mathcal V}_{i,j}^{(3)}(k,{\rm q}) = {\bf d}_{1,i}(k,{\rm q}) \cdot [{\bf m}_{1,j}(k,{\rm q}) \times {\bf d}_{j,i}(k,{\rm q})], 
\end{align}
\end{subequations}
with $i,j = 2 \rightarrow 5, j\neq i$, are scalar quantities, the triple products
\begin{subequations}
\begin{align}
& {\mathcal V}_{i,j}^{(1)}(k,{\rm q}) = {\bf d}_{1,i}(k,{\rm q}) \cdot [{\bf d}_{j,i}(k,{\rm q}) \times {\bf d}_{1,j}(k,{\rm q})], \\
& {\mathcal V}_{i,j}^{(4)}(k,{\rm q}) = {\bf d}_{1,i}(k,{\rm q}) \cdot [{\bf m}_{j,i}(k,{\rm q}) \times {\bf m}_{1,j}(k,{\rm q})],
\end{align}
\end{subequations}
are chirally-sensitive pseudoscalars [$\mathcal{V}_{i,j}^{(1)}(k,{\rm R})=-\mathcal{V}_{i,j}^{(1)}(k,{\rm S})$ and $\mathcal{V}_{i,j}^{(4)}(k,{\rm R})=-\mathcal{V}_{i,j}^{(4)}(k,{\rm S})$] representing effective chiral interaction volumes dependent over the orientation of electric/magnetic dipole matrix elements. Fig.~ \ref{Fig1}(d) illustrates the MD-induced temporal evolution of the electronic contribution to  $\mathcal{V}_{i,j}^{(1)}$ and $\mathcal{V}_{i,j}^{(4)}$, calculated for pure S enantiomer of reparixin in the conformation state $k = 1$. We incorporate these microscopic ingredients as ensemble-averaged quantities into our quantum-mechanical treatment to describe the macroscopic NL response of the chiral drug solution, based on the time averages of QMOs over $\simeq {\rm ns}$ MD timescales under the ergodic assumption.
The radiation-induced electron dynamics of reparixin, accounted for each conformation state $k=1,2,3,4$ and enantiomeric form ${\rm q = R,S}$, is governed by the density matrix equations
\begin{equation} \label{eq2}
    \dfrac{d\hat{\rho}_{k,{\rm q}}}{dt} = \dfrac{1}{i\hbar} \Big[ \hat{\mathcal{H}}_{\rm tot}^{(k,{\rm q})}, \hat{\rho}_{k,{\rm q}} \Big] + \hat{\mathcal L}(\hat{\rho}_{k,{\rm q}}),
\end{equation}
where $\hat{\rho}_{k,{\rm q}}({\bf r},t)$ denotes the time-dependent electronic density matrix and $\hat{\mathcal L}(\hat{\rho}_{k,{\rm q}})$ indicates the Lindblad operator, which includes the interaction with the thermal bath. The corresponding Lindblad relaxation rates for each electronic transition are retrieved from the bandwidth of the absorption spectra obtained from MD-PMM calculations \cite{Venturi2023}. For quasi-monochromatic external radiation fields ${\bf E}({\bf r},t) = {\rm Re} [{\bf E}_0 ({\bf r},t) e^{-i\omega t}]$ and ${\bf B}({\bf r},t) = {\rm Re} [{\bf B}_0 ({\bf r},t) e^{-i\omega t}]$ (such that $|\partial_t{\bf E}_0 ({\bf r},t)|<<\omega|{\bf E}_0 ({\bf r},t)|$, $|\partial_t{\bf B}_0 ({\bf r},t)|<<\omega|{\bf B}_0 ({\bf r},t)|$) with carrier wavelength $\lambda = 2 \pi c/ \omega$, where $c$ is the speed of light in vacuum, we perturbatively solve Eq.~\eqref{eq2} at second-order in the slowly-varying envelope approximation obtaining $\hat{\rho}_{k,{\rm q}}({\bf r},t)$. In order to do so, we express the density matrix $\hat{\rho}_{k,{\rm q}} = \sum_{i,j =1}^5 \rho_{i,j} |\phi_i(k,{\rm q}) \rangle \langle \phi_j(k,{\rm q})|$ in the basis set of one-electron eigenstates of the radiation-unperturbed Hamiltonian, where $\rho_{i,j}$ are the time-dependent density matrix elements. In the weak excitation limit, the multiple scale expansion of the density matrix elements is expressed as $\rho_{i,j} = \rho_{i,j}^{(0)} + a\rho_{i,j}^{(1)} + a^2\rho_{i,j}^{(2)} + ...,$ where $a$ indicates the perturbation strength up to second order, accounting for the diagonal $\rho_{1,1}^{(i \neq 1)} = 1 + \rho_{11}^{(2)}$, $\rho_{i,i}^{(i \neq 1)} = \rho_{ii}^{(2)}$ and off-diagonal $\rho_{1,i}^{(i \neq 1)} = \rho_{1i}^{(1)} + \rho_{1i}^{(2)}$ elements. Incorporating such expansions, we solve Eq.~\eqref{eq2}, obtaining the second-order contributions of the density matrix elements that correspond to the OR. The explicit expressions of such responses $\rho_{11}^{(2)} = \mathcal{P}_{11}^{(0)}, \rho_{ii}^{(2)} = \mathcal{P}_{ii}^{(0)}$, and $\rho_{1i}^{(2)} = \mathcal{P}_{1i}^{(0)}$ are given in the Supplementary Information (SI). In turn, for each enantiomeric form q = R, S and conformation $k=1,2,3,4$ of reparixin, we calculate the expectation value of the second-order OR electric dipole moment operator ${\bf d}_{\rm q}^{\rm (OR)} = \sum_{k=1}^4 p{(k)} {\rm Tr} \left[\hat{\rho}_{k,{\rm q}} \hat{\bf d}(k,{\rm q}) \right]= \sum_{k=1}^4 p{(k)}  \left[{\bf d}_{1,1} \mathcal{P}_{11}^{(0)} + \sum_{i=2}^5{\bf d}_{i,i} \mathcal{P}_{11}^{(0)} + \sum_{i=2}^5 {\rm Re}\left\{2{\bf d}_{i,1} \mathcal{P}_{1i}^{(0)} \right\} \right]$, accounting for the $p(k) \simeq 0.25$ probabilities of every thermodynamic realization $k$ for each enantiomeric form ${\rm q}$. In order to account for the random orientation of the reparixin molecules, we calculate averaged induced electric dipole moment over arbitrary rotations utilizing the Euler Rotation matrix $\hat{R}(\phi, \theta, \chi)$, where $\phi, \theta,$ and $\chi$ are the Euler angles. Thus, the rotational average of the induced electric dipole moments $\langle {\bf d}_{\rm q}^{\rm (OR)} \rangle$, see Eq.~\eqref{eq3}, for an isotropic chiral drug solution is calculated according to $\langle f \rangle = 1/(8\pi^2) \int_{0}^{2\pi} d\phi \int_{0}^{\pi} d\theta \int_{0}^{2\pi} d\chi f ({\phi, \theta, \chi}) \sin \theta$, where $f$ is an arbitrary vector quantity and the factor $\sin \theta$ accounts for the Haar-measure \cite{Andrews2004, Ohnoutek2022}, providing
\begin{equation} \label{eq3}
  \left\langle {\bf d}_{\rm q}^{\rm (OR)} \right\rangle = \epsilon_0 {\rm Re} \left[ \beta_{\rm EE}^{\rm (q)} ({\bf E}_0^* \times {\bf E}_0) + c \beta_{\rm EB}^{\rm (q)} ({\bf E}_0^* \times {\bf B}_0) + c^2 \beta_{\rm BB}^{\rm (q)} ({\bf B}_0^* \times {\bf B}_0) \right],
\end{equation}
where $\epsilon_0$ is the vacuum permittivity,  
\begin{subequations} \label{eq4}
    \begin{align}
    &\beta_{\rm EE}^{\rm (q)} (\omega) = \sum_k^4  p(k) \sum_{i=2}^5 \sum_{j=2, j\neq i}^5 \mathcal{G}_{i,j,k}^{(1)}(\omega) \mathcal{V}_{i,j}^{(1)} (k,{\rm q}), \\
    &\beta_{\rm EB}^{\rm (q)} (\omega) = \sum_k^4  p(k) \sum_{i=2}^5 \Big\{ \mathcal{G}_{i,k}(\omega) \mathcal{V}_{1i} (k,{\rm q}) + \sum_{j=2}^5 \Big[\mathcal{G}_{i,j,k}^{(2)}(\omega) \mathcal{V}_{i,j}^{(2)} (k,{\rm q}) + \mathcal{G}_{i,j,k}^{(3)}(\omega) \mathcal{V}_{i,j}^{(3)} (k,{\rm q}) \Big]\Big\},   \\
    &\beta_{\rm BB}^{\rm (q)} (\omega) = \sum_k^4  p(k) \sum_{i=2}^5 \sum_{j=2,j\neq i}^5 \mathcal{G}_{i,j,k}^{(4)}(\omega) \mathcal{V}_{i,j}^{(4)} (k,{\rm q}),
  \end{align}
\end{subequations}
are the NL-OR isotropic hyper-polarizabilities of each enantiomer ${\rm q=R,S}$, and $\mathcal{G}_{i,k}(\omega),\mathcal{G}_{i,j,k}^{(1-4)}(\omega)$ are wavelength-dependent coefficients reported in the SI. Note that, as anticipated, the hyper-polarizabilities depend over the triple products described above, accounting for the effective chiral interaction volume. In turn, while $\beta_{\rm EB}^{\rm (R)} (\omega) = \beta_{\rm EB}^{\rm (S)} (\omega)$ does not depend over the enantiomeric form, $\beta_{\rm EE}^{\rm (R)} (\omega)=-\beta_{\rm EE}^{\rm (S)} (\omega)$ and $\beta_{\rm BB}^{\rm (R)} (\omega)=-\beta_{\rm BB}^{\rm (S)} (\omega)$ are pseudoscalars. Note that, owing to the magnetic transition dipole moments, the isotropic chiral drug solutions acquire additional hyper-polarizabilities involving ${\bf E}_0^*\times{\bf B}_0$ and ${\bf B}_0^*\times{\bf B}_0$ cross products. Indeed, because Levi-Civita is the only tensor invariant over rotations in three dimensions, it mediates second-order NL interactions \cite{Bloembergen1969}. Figs.~\ref{Fig2}(a-c) illustrate the dependence of $\beta_{\rm EE}^{\rm (S)}$, $\beta_{\rm EB}^{\rm (S)}$, and $\beta_{\rm BB}^{\rm (S)}$ over the vacuum wavelength $\lambda$ of impinging radiation for pure S-reparixin enantiomer dissolved in water. We emphasize again that the QMOs (electric/magnetic dipole moments and electronic excitation energies) are incorporated as ensemble averages over the statistically relevant free energy basins probabilities \cite{Venturi2023}. Note that all the hyper-polarizabilities are resonant around $\lambda = 250 \ {\rm nm}$, where all the leading electronic transitions (broadened by collisions) considered in our calculations are convoluted in a single broad absorption peak.

\subsection{Macroscopic NL response of reparixin in water}

\begin{figure*}[t!]
\includegraphics[width=0.99\textwidth]{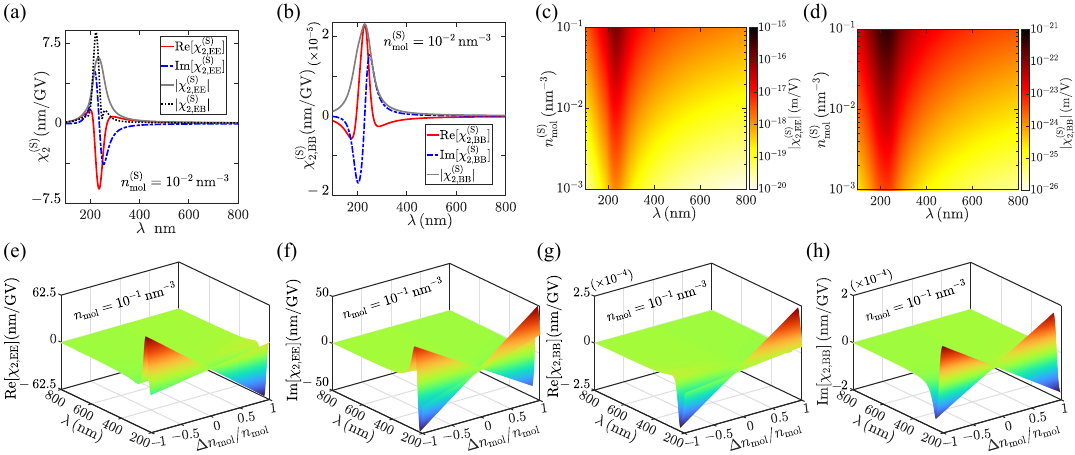}
\caption{ (\textbf{a},\textbf{b}) Dependence of (\textbf{a}) $\chi_{2, \rm EE}^{\rm(S)}$ and $\chi_{2, \rm EB}^{\rm (S)}$ and (\textbf{b}) $\chi_{2, \rm BB}^{\rm(S)}$ over the vacuum wavelength $\lambda$ at fixed molecular number density $n_{\rm mol} = n_{\rm mol}^{\rm (S)} = 10^{-2}\ {\rm nm}^{-3}$ of pure S-reparixin enantiomer dissolved in water. (\textbf{c},\textbf{d}) Color plots indicating  dependence of (\textbf{c}) $|\chi_{2, \rm EE}^{(\rm S)}|$ and (\textbf{d}) $|\chi_{2, \rm BB}^{\rm (S)}|$ over $\lambda$ and $n_{\rm mol}^{(\rm S)}$. Dependence of (\textbf{e}) ${\rm Re}[\chi_{2, \rm EE}]$, (\textbf{f}) ${\rm Im}[\chi_{2, \rm EE}]$, (\textbf{g}) ${\rm Re}[\chi_{2, \rm BB}]$, and (\textbf{h}) ${\rm Im}[\chi_{2, \rm BB}]$ over $\lambda$ and enantiomeric imbalance $\Delta n_{\rm mol} = n_{\rm mol}^{(\rm S)} - n_{\rm mol}^{(\rm R)}$ at fixed total molecular number density $n_{\rm mol}=10^{-1}\ {\rm nm}^{-3}$ of reparixin dissolved in water.}
\label{Fig3}
\end{figure*}

In the limit of dilute molecular density, the averaged macroscopic NL response of reparixin can be obtained by incorporating the second-order NL polarization ${\bf P}^{\rm (OR)} ({\bf r},t) = \sum_{\rm q = R,S} n_{\rm mol}^{(\rm q)} \left\langle {\bf d}_{\rm q}^{\rm (OR)} \right\rangle = {\rm Re} \left[{\bf P}_0^{\rm (OR)}({\bf r},t)\right]$ in the total polarization field. In turn, by introducing the displacement vector ${\bf D}({\bf r},t)= {\rm Re}[{\bf D}_0({\bf r},t)e^{-i\omega t}]$ and the magnetic field ${\bf H}({\bf r},t)= {\rm Re}[{\bf H}_0({\bf r},t)e^{-i\omega t}]$, one gets the constitutive relations 
\begin{subequations}
\begin{align}
& {\bf D}_0 = \epsilon_0[ \epsilon_{\rm r}(\omega) {\bf E}_0 -i\mu_0c\kappa(\omega) {\bf H}_0 ] + {\bf P}_0^{\rm (OR)}, \label{CEq1} \\
& {\bf B}_0 = \mu_0[ i[\kappa(\omega)/\mu_0c] {\bf E}_0 + \mu_{\rm r}(\omega) {\bf H}_0 ], \label{CEq2}
\end{align}
\end{subequations}
where $\mu_0$ is the vacuum permeability, $\epsilon_{\rm r}(\omega),\mu_{\rm r}(\omega),\kappa(\omega)$ are previously reported reparixin relative dielectric permittivity ($\epsilon_{\rm r}$), magnetic permeability ($\mu_{\rm r}$), and chiral parameter ($\kappa\propto\Delta n_{\rm mol}$) \cite{Venturi2023}, 
\begin{equation} \label{eq5}
    {\bf P}_0^{\rm (OR)} ({\bf r},t) = \epsilon_0 \Big[ \chi_{2, \rm EE}^{(\omega,-\omega)}({\bf E}_0^* \times {\bf E}_0) + c \chi_{2, \rm  EB}^{(\omega,-\omega)}({\bf E}_0^* \times {\bf B}_0) + c^2 \chi_{2, \rm BB}^{(\omega,-\omega)}({\bf B}_0^* \times {\bf B}_0 ) \Big],
\end{equation}
and $\chi^{(\omega,-\omega)}_{2, \rm EE/EB/BB} = \sum_{\rm q = R,S}n_{\rm mol}^{(\rm q)}\beta^{\rm (q)}_{\rm EE/EB/BB}$ are the second-order NL-OR susceptibilities of solvated reparixin. Remarkably, in spite of the random molecular orientation, the chiral drug solution acquires second-order quasi-static NL susceptibilities $\chi_{2}^{(\omega, -\omega)} = \sum_{k=1}^4 [ \chi_{2}^{(\omega, -\omega)}(k,{\rm R})+\chi_{2}^{(\omega, -\omega)}(k,{\rm S})]$ with enantioselective components $\chi_{2, \rm EE/BB}^{(\omega, -\omega)}\propto \Delta n_{\rm mol}$ that are proportional to the number density enantiomeric imbalance $\Delta n_{\rm mol} = n_{\rm mol}^{\rm (S)} - n_{\rm mol}^{\rm (R)}$  and a chirally-insensitive component $\chi_{2, \rm EB}^{(\omega, -\omega)}\propto n_{\rm mol}$ proportional to the total molecular number density $n_{\rm mol}=n_{\rm mol}^{\rm (S)} + n_{\rm mol}^{\rm (R)}$ of reparixin (irrespectively of the specific enantiomeric form content). Note that the resulting NL QSP ${\bf P}_0^{\rm (OR)}$, reminiscent of the inherent order imparted by the chirality of the isotropic assembly constituents, is always parallel to the driving radiation Poynting vector and spin density, which break the symmetry of the isotropic system. Figs.~\ref{Fig3}(a-d) depict the dependence of $\chi_{2, \rm EE}^{(\omega, -\omega)}$ and $\chi_{2, \rm EB}^{(\omega, -\omega)}$ over $\lambda$ and  $n_{\rm mol}$ for pure S-reparixin enantiomer ($n_{\rm mol}^{\rm (R)} = 0$) dissolved in water while, in Figs.~\ref{Fig3}(e-h), we report the dependence of such complex quantities over the normalized enantiomeric imbalance $\Delta n_{\rm mol}/ n_{\rm mol}$ and $\lambda$ for fixed $n_{\rm mol}=10^{-1} \ {\rm nm}^{-3}$. Note that, for dilute number densities $n_{\rm mol} = 10^{-3} - 10^{-1} \ {\rm nm^{-3}}$, the chirally-sensitive susceptibilities are of the order of $|\chi_{2, \rm EE}| \simeq 10^{-17} - 10^{-16} \ {\rm m/V}$, see Fig.~\ref{Fig3}(c), and $|\chi_{2,\rm BB}| \simeq 10^{-23} - 10^{-21} \ {\rm m/V}$, see Fig.~\ref{Fig3}(d), in the optical/near-infrared spectral range. In addition, for a racemic mixture where $n_{\rm mol}^{(\rm R)} = n_{\rm mol}^{(\rm S)}$, both ${\rm Re}[\chi_{2,\rm EE/BB}^{(\omega, -\omega)}]$ and ${\rm Im}[\chi_{2,\rm EE/BB}^{(\omega, -\omega)}]$ vanish, see Fig.~\ref{Fig3}(e-h). Consequently, the OR-induced voltage becomes polarization-insensitive for racemic mixtures, see below. 

\subsection{Optical rectification in PMCs embedding isotropic chiral samples}

In our proposed device, schematically depicted in Fig.~\ref{Fig1}(a), a layer of reparixin dissolved in water is placed at $0<z<d$ in between two ITO electrodes placed at $z \leq 0 \ \text{and} \ z \geq d$ with relative dielectric constant $\epsilon_{\rm t}(\omega)$ \cite{Landoni2014}. We investigate the electromagnetic excitation in the quasi-monochromatic approximation (QMA) by considering radiation with carrier wavelength $\lambda$ and arbitrary polarization expressed as a superposition of right $(s=+1)$ and left $(s=-1)$ circular polarization (CP) components, impinging on such a system from $z<0$. In the considered chiral medium, CP waves are eigenvectors of Maxwell's equations, and hence, we project the incident electric/magnetic fields onto the impinging CP unit vectors $\hat{\bf e}_s^{(+)}= (1/ \sqrt{2})(\hat{\bf e}_x +is \hat{\bf e}_y)$ and analyze the scattering dynamics by solving the macroscopic Maxwell’s equations, incorporating the chiral medium constitutive relations indicated in Eqs. (\ref{CEq1},\ref{CEq2}). In turn, we get
\begin{subequations}  \label{eq6}
    \begin{align}
    &{\bf E}({\bf r},t) = \sum\limits_{s=\pm 1} {\rm Re} \left\{ \left[ {\bf E}_{<}^{(s)} (z,t) \Theta(-z) + {\bf E}_{\rm IN}^{(s)} (z,t) \Theta_{\rm IN}(z) + {\bf E}_{>}^{(s)} (z,t) \Theta(z-d) \right] e^{-i\omega t} \right\}, \\
    &{\bf B}({\bf r},t) = \sum\limits_{s=\pm 1} {\rm Re} \left\{  \dfrac{1}{c} \left[ -i s \sqrt{\epsilon_t} {\bf E}_{<}^{(s)} (z,t) \Theta(-z) + i \left( \kappa - s \sqrt{\epsilon_{\rm r} \mu_{\rm r} } \right) {\bf E}_{\rm IN}^{(s)} (z,t) \Theta_{\rm IN}(z) + \right. \right. \nonumber \\
    & \hspace{3cm} \left. \left. -i s \sqrt{\epsilon_t} {\bf E}_{>}^{(s)} (z,t) \Theta(z-d) \right] e^{-i\omega t} \right\}, 
\end{align}
\end{subequations}
where $\Theta (z)$ is the Heaviside step function, $\Theta_{\rm IN} (z) = \Theta (z) - \Theta (z-d)$ and
\begin{subequations}  \label{eq:ch6_OR_allfields2}
    \begin{align}
    &{\bf E}_{<}^{(s)} =   E_{0,s}(z,t) \hat{\bf e}_s^{(+)} e^{ik_0 \sqrt{\epsilon_{\rm t}} z} + E_{{\rm R},s}(z,t) \hat{\bf e}_s^{\rm (-)} e^{-ik_0 \sqrt{\epsilon_{\rm t}} z}, \\
    &{\bf E}_{\rm in}^{(s)} ({\bf r}) = \sum\limits_{\sigma = \pm 1} E_{s}^{(\sigma)}(z,t) \hat {\bf e}^{(\sigma)}_s e^{i \sigma \beta_s z}, \\
     &{\bf E}_{>}^{(s)} ({\bf r}) =   \sum\limits_{\sigma = \pm 1} E_{{\rm T},s}(z,t) \hat {\bf e}^{({\rm +})}_s e^{i k_0 \sqrt{\epsilon_2} z},
\end{align}
\end{subequations}
$\omega = 2 \pi c/ \lambda$ indicates the angular frequency, $k_0 = \omega /c$, $E_{0,\pm 1}(z,t)$ represent the projections of the impinging vectorial envelopes (weakly depending over $z,t$ in the QMA, justified by the considered $t_0 \simeq$ ns pulse duration, implying a uniform envelope over $ct_0\simeq 30$ cm $>>d=100$ $\mu$m) onto $\hat{\bf e}^{(+)}_{\pm 1}$, $\hat{\bf e}_s^{\rm (\sigma)} = (1/ \sqrt{2})(\sigma\hat{\bf e}_x +is \hat{\bf e}_y)$ are the forward ($\sigma = +1$) and backward ($\sigma = -1$) CP unit vectors, and $\beta_{s} (\lambda) = k_0[\sqrt{\epsilon_{\rm r}(\lambda) \mu_{\rm r}(\lambda)} + s\kappa(\lambda)]$ is the polarization-dependent wavenumber within the chiral medium. The unknown CP amplitudes $E_{ {\rm R/T},s}$ and $E_{s}^{(\sigma)}$ are obtained analytically in the QMA by imposing the boundary conditions for the continuity of (i) the normal components of the displacement vector and induction magnetic field, and (ii) the tangential components of the electric and magnetic fields at the interfaces $z=0, d$. We provide the cumbersome analytical expressions for such field amplitudes in the SI. In order to calculate the OR-induced voltage at the ITO boundaries, we insert Eqs. (\ref{eq6}) in the NL QSP given by Eq. (\ref{eq5}) and analytically evaluate the OR-induced voltage difference burst $\Delta V(t)$ between the two ITO-electrodes, obtaining  
\begin{align} \label{eq:ch6_OR_Volt_final}
   \Delta V (t) &=  \dfrac{1}{2\epsilon_{\rm S}}\sum_{s=\pm 1} {\rm Re} \left\{ s \left[ i \chi_{2, \rm EE}^{(\omega,-\omega)} + \zeta_s^*(\lambda) \chi_{2, \rm EB}^{(\omega,-\omega)} + |\zeta_s (\lambda) |^2 \chi_{2, \rm BB}^{(\omega,-\omega)} \right]  \left[ \mathcal{A}_1(\lambda,t) d + \mathcal{A}_2(\lambda,t) \right] \right\} + \nonumber \\
    &+\dfrac{1}{2\epsilon_{\rm S}}\sum_{s=\pm 1} \sum_{s^{\prime}=\pm 1} \left\{ s \left[ i\chi_{2, \rm EE}^{(\omega,-\omega)} +  \zeta_{s^{\prime}}^*(\lambda) \chi_{2, \rm EB}^{(\omega,-\omega)} +i \zeta_s(\lambda) \zeta_{s^{\prime}}^*(\lambda) \chi_{2, \rm BB}^{(\omega,-\omega)} \right] \left[ \mathcal{A}_3(\lambda,t) d + \mathcal{A}_4(\lambda,t) \right] \right\},
\end{align}
where $\epsilon_{\rm S}\simeq 78.5$ is the water static relative dielectric permittivity at ambient temperature, $\zeta_s (\lambda) = -s\beta_{-s}/k_0$, and $\mathcal{A}_{1-4}(\lambda,t)$ are involved functions, explicitly given in the SI, containing field amplitudes and depending parametrically over time through the impinging envelopes $E_{0,\pm}(0,t)=A_0 e^{-t^2/4t_0^2}$. Note that $\Delta V (t)$ contains both chirally-insensitive contributions arising from $\chi_{2, \rm EB}^{(\omega,-\omega)}$, leading to the background voltage 
\begin{equation}
\Delta V_{\rm bckg}(t) = \frac{1}{2\epsilon_{\rm S}} \sum_{s=\pm 1} {\rm Re} \left\{ s \left[ \zeta_s^* \chi_{2, \rm EB}^{(\omega,-\omega)} \right] \left( \mathcal{A}_1 d + \mathcal{A}_2 \right) \right\} + \frac{1}{2\epsilon_{\rm S}} \sum_{s=\pm 1} \sum_{s^{\prime}=\pm 1} {\rm Re} \left\{ s \left[ \zeta_{s^{\prime}}^* \chi_{2, \rm EB}^{(\omega,-\omega)} \right] \left( \mathcal{A}_3 d + \mathcal{A}_4 \right) \right\},
\end{equation}
and an enantio-selective OR-voltage modulation $[\Delta V(t) - \Delta V_{\rm bckg}(t)] \propto \Delta n_{\rm mol}$ arising from the $\chi_{2, \rm EE/EB}^{(\omega,-\omega)}$ pseudoscalar terms in Eq. (\ref{eq:ch6_OR_Volt_final}). We determine $\Delta V(t)$ in realistic excitation conditions by launching an input Gaussian laser pulse with intensity temporal profile $I(t) = I_0 e^{-t^2/2t_0^2}$ onto the PMC, where $I_0=(1/2)\epsilon_0c|\tilde{A}_0|^2$ is the impinging peak intensity and $t_0$ is the pulse duration. Note that, in order to obtain realistic results, the input amplitude at the ITO-solvated reparixin boundary $A_0$ is determined from the impinging external amplitude $\tilde{A}_0$ (we assume air as external medium before the ITO electrode) through the Fresnel coefficient $A_0=2\tilde{A}_0/(1+{\rm Re}\sqrt{\epsilon_{\rm t}})e^{-k_0{\rm Im}\sqrt{\epsilon_{\rm t}}w}$, accounting for both reflection and absorption of the ITO-based electrode with thickness assumed to $w=10$ $\mu$m.     

\begin{figure*}[t!]
\includegraphics[width=0.99\textwidth]{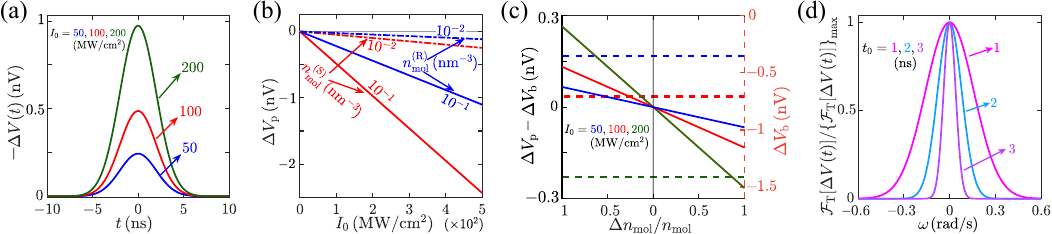}
\caption{(\textbf{a}) OR-induced time-dependent voltage burst $\Delta V(t)$ in the considered PMC with thickness $d=100 \ \mu{\rm m}$, see Fig.~\ref{Fig1}, for several impinging laser peak intensities $I_0=50,100,200$ MW$/$cm$^2$, fixed pulse duration $t_0 = 2$ ns, right (s=+1) impinging CP, fixed carrier vacuum wavelength $\lambda_{\rm p}=800 \ {\rm nm}$, calculated for pure S-reparixin enantiomers with number molecular density $n_{\rm mol}= n_{\rm mol}^{\rm (S)} = 10^{-1} \ {\rm nm}^{-3}$ dissolved in water. (\textbf{b}) Dependence of OR-induced voltage peak $\Delta V_{\rm p}$ over $I_0$ for pure S (red lines, $n_{\rm mol}=n_{\rm mol}^{\rm (S)}$) and pure R (blue lines, $n_{\rm mol}=n_{\rm mol}^{\rm (R)}$) reparixin enantiomers with $n_{\rm mol}=10^{-1}$ nm$^{-3}$ (full curves) and $n_{\rm mol}=10^{-2}$ nm$^{-3}$ (dashed lines), and other parameters coinciding with the ones in ({\bf a}). (\textbf{c}) Dependence of (left y-axis) $\Delta V_{\rm p}-\Delta V_{\rm b}$ and (right y-axis) $\Delta V_{\rm b}$ over the enantiomeric excess density $\Delta n_{\rm mol} = n_{\rm mol}^{\rm (S)} - n_{\rm mol}^{\rm (R)}$ for several distinct impinging intensities $I_0=50,100,200$ MW$/$cm$^2$ and fixed total molecular number density $n_{\rm mol} = n_{\rm mol}^{\rm (S)} + n_{\rm mol}^{\rm (R)}= 10^{-1} \ {\rm nm}^{-3}$ of aqueous reparixin. (\textbf{d}) Normalized Fourier transform of output voltage signal $\mathcal{F}_{\rm T}[\Delta V(t)]/\{\mathcal{F}_{\rm T}[\Delta V(t)]\}_{\rm max}$ as a function of the frequency $\nu=\omega/2\pi$ for several impinging pulse durations $t_0=1,2,4$ ns at fixed $I_0 = 100~{\rm MW/cm^2}$, obtained for pure S-reparixin enantiomer with $n_{\rm mol}=n_{\rm mol}^{(\rm S)}$ dissolved in water.}
\label{Fig4}
\end{figure*}

\section{Discussion}

\begin{figure*}[t!]
\includegraphics[width=0.5\textwidth]{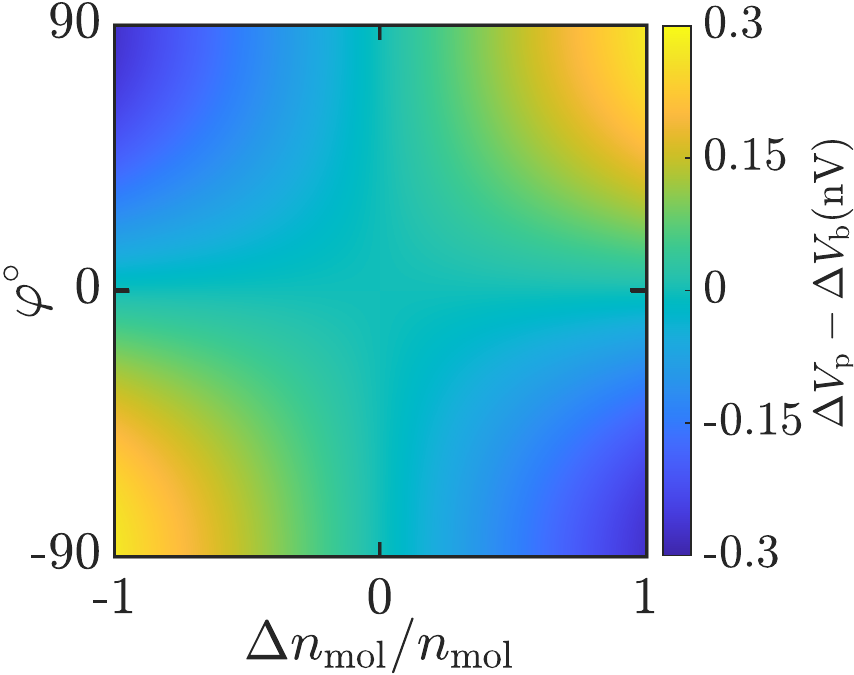}
\caption{Dependence of $\Delta V_{\rm p}-\Delta V_{\rm b}$ over the excitation ellipticity $\varphi$ and the normalized enantiomeric imbalance $\Delta n_{\rm mol}/n_{\rm mol}$ for fixed reparixin total number density $n_{\rm mol}=10^{-1}$ nm$^{-3}$, and impinging radiation carrier wavelength $\lambda=800$ nm and peak intensity $I_0 = 200$ MW$/$cm$^2$.}
\label{Fig5}
\end{figure*}

The calculated voltage burst $\Delta V(t)$ is reported in Fig.~\ref{Fig4}(a), indicating the dependence of $\Delta V(t)$ over time $t$ at fixed impinging vacuum wavelength $\lambda = 800 \ {\rm nm}$ and pulse duration $t_0 = 2 \ {\rm ns}$ for right $(s=+1)$ CP impinging radiation with several peak intensities and fixed molecular number density $n_{\rm mol} = n_{\rm mol}^{\rm (S)} = 10^{-1} \ {\rm nm}^{-3}$ of pure S-reparixin enantiomers dissolved in water. Note that, in such realistic conditions, for intensities $I_0 \approx \ {\rm MW/cm^2}$, one can achieve an efficient OR-induced voltage burst with peak $\Delta V_{\rm p} \approx \ {\rm nV}$ between the ITO electrodes. In Fig.~\ref{Fig4}(b), we illustrate the dependence of $\Delta V_{\rm p}$ over $I_0$ at fixed $\lambda = 800 \ {\rm nm}$ for two distinct molecular densities, labelled by full and dashed lines, of pure ${\rm R}$ (blue lines, $n_{\rm mol}=n_{\rm mol}^{\rm (R)}$) and pure ${\rm S}$ (red lines, $n_{\rm mol}=n_{\rm mol}^{\rm (S)}$) reparixin enantiomers dissolved in water. Fig.~\ref{Fig4}(c) illustrates the dependence of (left y-axis) $\Delta V_{\rm p}-\Delta V_{\rm b}$ and (right y-axis) $\Delta V_{\rm b}={\rm max}[\Delta V_{\rm bckg}(t)]$ over the enantiomeric excess density $\Delta n_{\rm mol}/n_{\rm mol}$ for several impinging intensities $I_0$ at fixed total molecular number density $n_{\rm mol} = 10^{-1}~{\rm nm}^{-3}$. Note that, as anticipated, the enantioselective signal $\Delta V_{\rm p}-\Delta V_{\rm b}$ is sitting on top of a constant background voltage $\Delta V_{\rm b}$ which can be isolated thanks to its invariance over enantiomeric imbalance. We emphasize that $\Delta V_{\rm b}$ depends over the total number density of chiral molecules $n_{\rm mol}$ irrespectively of their enantiomeric form (it adds up for ${\rm R}$ and ${\rm S}$ enantiomers), and it would not be generated if the molecules were not chiral. Also note that for racemic chiral mixtures $\Delta V_{\rm p}$ vanishes and flips sign depending on the enantiomeric excess of the chiral mixture. In order to determine the spectral response of the output voltage signal, we calculate the Fourier transform (FT) $\mathcal{F}_{\rm T}\Delta V(t)$ of the induced voltage. Fig.~\ref{Fig4}(d) depicts the variation of the normalized FT $\mathcal{F}_{\rm T}\Delta V(t)/[\mathcal{F}_{\rm T}\Delta V(t)]_{\rm max}$ over the frequency $\nu=\omega/2\pi$ for several impinging pulse durations $t_0$ (the pulse full-width at half-maximum is FWHM=$2\sqrt{2{\rm ln}2}t_0$) and fixed $I_0 = 100\ {\rm MW/cm^2}$, calculated for fixed $n_{\rm mol} = 10^{-1}~{\rm nm}^{-3}$ of pure S-reparixin enantiomer dissolved in water. Note that the bandwidth of such a voltage signal increases significantly when the pulse duration of the impinging laser decreases.
Note that the enantioselective signal $[\Delta V_{\rm p}-\Delta V_{\rm b}] \propto {\rm sin}\varphi$ is proportional to ${\rm sin}\varphi$, where $\varphi$ is the impinging radiation ellipticity, defined through the impinging electric field vectorial envelope ${\bf E}_0(z,t) = A_0(z,t)\left(\hat{\bf e}_x + e^{i\varphi}\hat{\bf e}_y\right)/\sqrt{2}$ with amplitude $A_0(z,t)$. In turn, $\Delta V_{\rm p}-\Delta V_{\rm b}$ vanishes for linear polarization excitation and flips sign for opposite impinging CPs, see Fig. \ref{Fig5}, where we illustrate the dependence of $\Delta V_{\rm p}-\Delta V_{\rm b}$ over the excitation ellipticity $\varphi$ and the normalized enantiomeric imbalance $\Delta n_{\rm mol}/n_{\rm mol}$ for fixed reparixin total number density $n_{\rm mol}=10^{-1}$ nm$^{-3}$, and impinging radiation carrier wavelength $\lambda=800$ nm and peak intensity $I_0 = 200$ MW$/$cm$^2$.

From a practical perspective, by assuming an ITO-based flat capacitor with plate lateral dimensions $l\simeq 500$ $\mu$m separated by the distance $d = 100$ $\mu$m considered in our calculations, the interaction volume of the considered chiral drug is $V_{\rm int} \simeq 25$ nl. Note that, by reducing the plates distance to $d \simeq 10$ $\mu$m, one gets $V_{\rm int} \simeq$ nl. However, reducing the thickness implies an increase of the device capacitance $C=\epsilon_0\epsilon_{\rm S}l^2/d$, producing increased relaxation time $\tau = RC\simeq 2$ ns when the system is connected to an amperometer and an amplification system with total resistance assumed to $R\simeq 100$ $\Omega$. In turn, by reducing the capacitor spacing the $RC$ transient integrates the temporal profile of the enantioselective signal complicating the measurement. For this reason, the reported results are obtained for larger spacing $d=100$ $\mu$m. We finally emphasize that the predicted voltage bursts can be measured experimentally by employing a lock-in amplifier \cite{DeMarcellis2007, Kishore2020} enabling to retrieve the enantiomeric excess density of the considered chiral drug solution by measuring the OR-induced voltage.

\section{Conclusions}

In this work, we devise a novel chiroptical sensing approach capable of measuring the enantiomeric excess of chiral mixtures with $\simeq 20$ nl volume. The proposed innovative technique is based on the hitherto unknown OR effect of isotropic chiral mixtures, which we model from first principles in the electric/magnetic dipole approximation calculating molecular hyper-polarizabilities and deriving the macroscopic NL response. Furthermore, we devise PMCs embedding the dilute chiral drug solutions to be analyzed, specifying our calculations to reparixin, an inhibitor of the CXCR2 function attenuating inflammatory responses that has been adopted in clinical trials for the treatment of hospitalized patients with COVID-19 pneumonia. Our results demonstrate that, upon CP optical excitation of the PMC, chirality detection is enabled by the measurement of output voltage signal with peak $\simeq$ nV. We emphasize that the electrical detection feature of our proposed optoelectronic device is appealing for lab-on-a-chip and microfluidic integration, paving the way for advanced microdevices capable to detect chirality of nl samples, with applications in nanomedicine and drug discovery.

\section{Acknowledgments}

This work has been partially funded by the European Union - NextGenerationEU under the Italian Ministry of University and Research (MUR) National Innovation Ecosystem grant ECS00000041 - VITALITY - CUP E13C22001060006. This work has been supported by the European Union under grant agreement No 101046424. Views and opinions expressed are however those of the author(s) only and do not necessarily reflect those of the European Union or the European Innovation Council. Neither the European Union nor the European Innovation Council can be held responsible for them. J.B. additionally acknowledges financial support from the European Research Council for ERC Proof of Concept Grant (101248732), Lasers4EU (HEU-GA 101131771) , Programa Estatal de Transferencia y Colaboración, Agencia Estatal de Investigación, PDC2025-165773-I00 / financiado por MICIU/AEI / 10.13039/501100011033, “QLOGIC - Driving Digital Transformation with Quantum Innovation” was supported by the Secretariat of Digital Policies of the Government of Catalonia (G.A. GOV/51/2022), the Spanish Ministry of Economy and Competitiveness, through the “Severo Ochoa” Center of Excellence (CEX2024-001490-S [MICIU/AEI/10.13039/501100011033], Project PID2024-162757NB-I00 funded by MICIU/ AEI / 10.13039/501100011033/ FEDER, UE, the Catalan Agencia de Gestió d’Ajuts Universitaris i de Recerca (AGAUR) (SGR-2021-01449), Fundació Cellex Barcelona and the CERCA Programme / Generalitat de Catalunya.", and the Spanish State Research Agency (AEI) for PID2024-162757NB-I00 (AMAPOLA) funded by MCIN/AEI /10.13039/501100011033, RED2024-154107-E.

The authors acknowledge fruitful discussions with Giovanna Salvitti, Andrea Toma, Francesco Di Stasio, Francesco Tani, Hatice Altug, Patrice Genevet, Samira Khadir, Remi Colom, Michele Dipalo, Giovanni Melle, Sotirios Christodoulou, and Anna Maria Cimini.

\section{Author contributions}

A.M. conceived the idea. R.A., A.S., M.V., M.A., and A.Mar. developed the theoretical analysis and numerical calculations. L.A., A.D.M., A.Mec., J.B., D.T, and C.F. discussed the results and proposed experimentally viable materials and setups. This work was supervised by A.Mar. R.A. and A.Mar. prepared the paper with input from all authors.

\section{SI: Solution of density matrix equations}

The density matrix equations solution at second order, see main text, contains the following parameters
\begin{align}
    &\mathcal{P}_{11}^{(0)} = -\sum_{i=2}^N{\rm Im} \left\{ \dfrac{ i|{\mathbf d}_{i,1} \cdot {\mathbf E}_0 + {\mathbf m}_{i,1} \cdot {\mathbf B}_0 |^2}{\gamma_i \hbar^2 \left\{ \gamma_i + 2i  \left[ \omega - \left( \omega_i - \omega_1 \right) \right] \right\}  }  + \dfrac{ i| {\mathbf d}_{1,i} \cdot {\mathbf E}_0 + {\mathbf m}_{1,i} \cdot {\mathbf B}_0 |^2}{\gamma_i \hbar^2  \left\{ \gamma_i - 2i \left[ \omega +\left( \omega_i - \omega_1 \right) \right] \right\}  } \right\}, \\
&\mathcal{P}_{ii}^{(0)} = {\rm Im} \left\{ \dfrac{ i|{\mathbf d}_{i,1} \cdot  {\mathbf E}_0 + {\mathbf m}_{i,1} \cdot  {\mathbf B}_0  |^2}{\gamma_i \hbar^2 \{ \gamma_i + 2i  [ \omega - (\, \omega_i - \omega_1 )\, ] \} } +  \dfrac{ i|{\mathbf d}_{1,i} \cdot  {\mathbf E}_0 + {\mathbf m}_{1,i} \cdot  {\mathbf B}_0 |^2}{\gamma_i \hbar^2  \{ \gamma_i-2i [ \omega+(\, \omega_i - \omega_1)\, ] \}  } \right\},\\
&\mathcal{P}_{1i}^{(0)}=
\left\{
\begin{aligned}
      &\dfrac{(\,{\mathbf d}_{1,i} \cdot  {\mathbf E}_0^* + {\mathbf m}_{1,i} \cdot  {\mathbf B}_0^*)\, [ (\, {\mathbf d}_{1,1} - {\mathbf d}_{i,i})\, \cdot  {\mathbf E}_0 + (\,{\mathbf m}_{1,1} - {\mathbf m}_{i,i})\, \cdot  {\mathbf B}_0 ] }{\hbar^2 [\, \gamma_i -2i (\, \omega_i- \omega_1)\,]\, \{\gamma_i + 2i [ \omega -(\, \omega_i-\omega_1)\, ] \} } 
     \\
     &+\dfrac{ (\, {\mathbf d}_{1,i} \cdot  {\mathbf E}_0 + {\mathbf m}_{1,i} \cdot  {\mathbf B}_0)\, [ (\,{\mathbf d}_{1,1} - {\mathbf d}_{i,i} )\, \cdot  {\mathbf E}_0^* + (\,{\mathbf m}_{1,1} - {\mathbf m}_{i,i})\,\cdot  {\mathbf B}_0^* ] }{\hbar^2 [ \gamma_i - 2i (\, \omega_i-\omega_1)\, ] \{\gamma_i - 2i [ \omega + (\, \omega_i-\omega_1)\, ] \} }   
    \\
    &-\sum_{\substack{n=2 \\ n \neq i}}^N \dfrac{ (\, {\mathbf d}_{1,n} \cdot  {\mathbf E}_0^* + {\mathbf m}_{1,n} \cdot  {\mathbf B}_0^*)\, (\,{\mathbf d}_{n,i} \cdot  {\mathbf E}_0 + {\mathbf m}_{n,i} \cdot  {\mathbf B}_0)\, }{ \hbar^2 \left[  \gamma_i-2i(\, \omega_i -\omega_1)\, \right] \{\gamma_n + 2i \left[ \omega-(\,\omega_n-\omega_1)\, \right] \}}  
    \\
    &-\sum_{\substack{n=2 \\ n \neq i}}^N \dfrac{(\,{\mathbf d}_{1,n} \cdot  {\mathbf E}_0 + {\mathbf m}_{1,n} \cdot  {\mathbf B}_0)\, (\,{\mathbf d}_{n,i} \cdot  {\mathbf E}_0^* + {\mathbf m}_{n,i} \cdot  {\mathbf B}_0^*)\, }{ \hbar^2 [ \gamma_i - 2i (\, \omega_i -\omega_1)\, ] \{\gamma_n - 2i [ \omega+(\,\omega_n-\omega_1) ] \}}.  
 \end{aligned}   
\right\},
\end{align}

\section{SI: Wavelength dependent coefficients of hyper-polarizabilities} 

Upon interaction with external radiation, the chiral drug solutions possess microscopic hyperpolarizabilities $\beta_{\rm EE}^{(q)}$, $\beta_{\rm EB}^{(q)}$, and $\beta_{\rm BB}^{(q)}$ characterized by pseudoscalar quantities combined with wavelength-dependent coefficients that are given by

\begin{subequations} \label{Seq1}
    \begin{align}
        \mathcal{G}_{i,j,k}^{(1)}(\omega) &= \dfrac{4i\omega}{ 3\hbar^2[\gamma_i - 2i \Delta \omega_{i}](\gamma_j + 2i[\omega - \Delta \omega_{j}])(\gamma_j - 2i[\omega + \Delta \omega_{j}]) },  \\
        \mathcal{G}_{i,j,k}^{(2)}(\omega) &=\dfrac{ -4i[\gamma_i \Delta \omega_j+\Delta \omega_i(\gamma_j - 2i \omega)] }{ 3\hbar^2[\gamma_i^2 +4 \Delta \omega_{i}^2](\gamma_j - 2i[\omega - \Delta \omega_{j}])(\gamma_j - 2i[\omega + \Delta \omega_{j}]) } , \\ 
        \mathcal{G}_{i,j,k}^{(3)}(\omega) &=\dfrac{4i[\gamma_i \Delta \omega_j + \Delta \omega_i(\gamma_j + 2i \omega)]}{ 3\hbar^2[\gamma_i^2 + 4 \Delta \omega_{i}^2](\gamma_j + 2i[\omega - \Delta \omega_{j}])(\gamma_j + 2i[\omega + \Delta \omega_{j}]) }  ,\\ 
        \mathcal{G}_{i,j,k}^{(4)}(\omega) &= \dfrac{4i\omega}{ 3\hbar^2[\gamma_i - 2i \Delta \omega_{i}](\gamma_j + 2i[\omega - \Delta \omega_{j}])(\gamma_j - 2i[\omega + \Delta \omega_{j}]) }  ,  \\
        \mathcal{G}_{i,k}(\omega) &= \dfrac{16\omega \Delta \omega_i}{3\hbar^2[\gamma_i^2 + 4 (\omega-\Delta \omega_{i})^2][\gamma_i^2 + 4 (\omega+\Delta \omega_{i})^2]} ,\nonumber\\
        &+ \dfrac{4i\Delta \omega_i(2\gamma_i + i\omega)}{ 3\hbar^2[\gamma_i^2 + 4 \Delta \omega_{i}^2](\gamma_i + 2i[\omega - \Delta \omega_{i}])(\gamma_i + 2i[\omega + \Delta \omega_{i}]) } ,
    \end{align}
\end{subequations}
where $\Delta \omega_i = \omega_i - \omega_1$ and $\Delta \omega_j = \omega_j - \omega_1$. The electronic relaxation rates $\gamma_i$ are extracted form the frequency bandwidth of the UV absorption spectrum and the electronic transition energies $\hbar \Delta \omega$  are calculated through MD-PMM simulations, see main text. Other parameters are defined in the main text.

\section{SI: Field amplitudes in the consider micro-cavity} 
In order to calculate the forward $(\sigma=+1)$ and backward $(\sigma=+1)$ field amplitude $E_s^{(\sigma)}$ within the chiral medium and the reflected/transmitted amplitudes $E_{ {\rm R/T},s}$, we impose the  boundary conditions for the continuity of (i) the normal components of the displacement vector and induction magnetic field, and (ii) the tangential components of the electric and magnetic fields at the interfaces $z=0,d$, obtaining
\begin{subequations}  \label{Seq2}
    \begin{align}
        &E_{{\rm R},-} = \left[ \left\{ { (1+u_1)(1-u_2)e^{i \beta_+ d} - (1-u_1)(1+u_2)e^{-i \beta_- d} }\right\} /Z_1 \right] E_{0,+} \\
        &E_{+}^{(+)} = \left[ \left\{  { 2(1+u_2)e^{-i \beta_- d} } \right\} /Z_1 \right] E_{0,+} \\
        &E_{-}^{(-)} = \left[ \left\{ { 2(1-u_2)e^{i \beta_+ d} } \right\} /Z_1 \right]  E_{0,+} \\
        &E_{ {\rm T},+} = \left[ \left\{ { 4 u_2e^{i (\beta_+ - \beta_-)  d -ik_0 \sqrt{\epsilon_2}d} }  \right\} /Z_1 \right]  E_{0,+} \\
        &E_{{\rm R},+} = \left[ \left\{ { -(1-u_1)(1+u_2)e^{-i \beta_+ d} + (1-u_2)(1+u_1)e^{i \beta_- d} } \right\} /Z_2 \right]    E_{0,-} \\
        &E_{-}^{(+)} = \left[ \left\{ { 2(1+u_2)e^{-i \beta_+ d} } \right\} /Z_2 \right] E_{0,-} \\
        &E_{+}^{(-)} = \left[ \left\{ { 2(1-u_2)e^{i \beta_- d} } \right\} /Z_2 \right] E_{0,-} \\
        &E_{ {\rm T},-} = \left[ \left\{  { 4 u_2e^{-i (\beta_+ - \beta_-)  d - i k_0 \sqrt{\epsilon_2} d } } \right\} /Z_2 \right] E_{0,-} ,
    \end{align}
\end{subequations}
where $u_1 = \sqrt{\epsilon_{\rm r}/ \mu_{\rm r} \epsilon_1}, u_2 = \sqrt{\epsilon_{\rm r}/ \mu_{\rm r} \epsilon_2}$ and
\begin{subequations} \label{Seq3}
    \begin{align}
        &Z_1 = { -(1 - u_1)(1-u_2)e^{i \beta_+ d} + (1+u_1)(1+u_2)e^{-i \beta_- d} }\\
        &Z_2={ (1 + u_2)(1 + u_1) e^{-i \beta_+ d} - (1 - z_2)(1 - u_1)e^{i \beta_- d} }.
    \end{align}
\end{subequations}
All parameters are defined in the main text.

\section{SI: Coefficients containing field amplitudes in the OR-induced voltage.} 

The OR-induced voltage difference $\Delta V$ between 
two ITO-electrodes contains field amplitudes dependent functions, see main text, that are explicitly given by

\begin{subequations} \label{Seq4}
    \begin{align}
    &\mathcal{A}_1(\lambda,t) = \left[ 
         \begin{aligned}
         &-|E_s^{(+)}(t)|^2 \left( 1 + e^{-2{\rm Im}[\beta_S]d} \right) - {E_s^{(-)}(t) }^* E_s^{(+)}(t) \left(1+e^{2i{\rm Re}[\beta_S]d} \right) \\
         &+ {E_s^{(+)}(t) }^* E_s^{(-)}(t) \left( 1 + e^{-2i{\rm Re}[\beta_S]d} \right) +  |E_s^{(-)}(t)|^2 \left( 1 + e^{2{\rm Im}[\beta_S]d} \right) 
         \end{aligned} \right], \\
    &\mathcal{A}_2(\lambda,t) = \left[ 
         \begin{aligned}
         &\dfrac{|E_s^{(+)}(t)|^2}{2{\rm Im}[\beta_S]}  \left(e^{-2{\rm Im}[\beta_s]d} - 1 \right) -\dfrac{{E_s^{(-)}(t)}^* E_s^{(+)}(t) }{2i{\rm Re}[\beta_S]} \left(e^{2i{\rm Re}[\beta_s]d} - 1 \right) \\
         & -\dfrac{{E_s^{(+)}(t)}^* E_s^{(-)}(t) }{2i{\rm Re}[\beta_S]} \left(e^{-2i{\rm Re}[\beta_s]d} - 1 \right) +\dfrac{|E_s^{(-)}(t)|^2}{2{\rm Im}[\beta_S]}  \left(e^{2{\rm Im}[\beta_s]d} - 1 \right)
         \end{aligned} \right], \\
    &\mathcal{A}_3(\lambda,t) = \left[ {E_{s^{\prime}}^{(-)}(t) }^* E_s^{(+)}(t) \left( 1 + e^{i \left(\beta_S+\beta_{S^{\prime} }^*\right)d} \right) -{E_{s^{\prime}}^{(+)}(t) }^* E_s^{(-)}(t) \left( 1 + e^{-i\left(\beta_S+\beta_{S^{\prime} }^* \right)d} \right)   \right], \\
    &\mathcal{A}_4(\lambda,t) = \left[ \dfrac{{E_{s^{\prime}}^{(-)}(t) }^* E_s^{(+)}(t)}{i(\beta_s+\beta_{s^{\prime}}^*)} \left(e^{ i(\beta_s+\beta_{s^{\prime}}^*)d } - 1 \right) +  \dfrac{{E_{s^{\prime}}^{(+)}(t) }^* E_s^{(-)}(t)}{i(\beta_s+\beta_{s^{\prime}}^*)} \left(e^{- i(\beta_s+\beta_{s^{\prime}}^*)d } - 1 \right)   \right].      
    \end{align}
\end{subequations}
All parameters are defined in the main text.

\end{document}